\documentclass[aps,twocolumn,prl,showpacs,superscriptaddress,groupedaddress]{revtex4}  
\usepackage{dcolumn}
\usepackage{bm}
\usepackage{color}

\def\ltape{\hbox{\ $<$\hskip -8pt\raise -4pt\hbox{$\sim$}\ }}
\def\gtape{\hbox{\ $>$\hskip -8pt\raise -4pt\hbox{$\sim$}\ }}

\pdfoutput=1

   \ifx\pdfoutput\undefined
   \usepackage[dvips]{graphicx}
   \else
   \usepackage[pdftex]{graphicx}
   \fi
   \usepackage{epstopdf}
   \DeclareGraphicsExtensions{.pdf,.gif,.jpg,.eps,.png}
  
   \usepackage{amsmath}
  \usepackage{amssymb}

\begin{document}

\title{Why does steady-state magnetic reconnection have
  a maximum local rate of order 0.1?}

\author{Yi-Hsin~Liu}
\affiliation{NASA-Goddard Space Flight Center, Greenbelt, MD 20771}
\author{M.~Hesse}
\affiliation{NASA-Goddard Space Flight Center, Greenbelt, MD 20771}
\author{F.~Guo}
\affiliation{Los Alamos National Laboratory, Los Alamos, NM 87545}
\author{W.~Daughton}
\affiliation{Los Alamos National Laboratory, Los Alamos, NM 87545}
\author{H.~Li}
\affiliation{Los Alamos National Laboratory, Los Alamos, NM 87545}
\author{P. A.~Cassak}
\affiliation{West Virginia University, Morgantown, WV 26506}
\author{M. A.~Shay}
\affiliation{University of Delaware, Newark, DE 19716}
\date{\today}

\begin{abstract}


Simulations suggest collisionless steady-state magnetic reconnection of Harris-type current sheets proceeds with a rate of order 0.1, independent of dissipation mechanism. We argue this long-standing puzzle is a result of constraints at the magnetohydrodynamic (MHD) scale. We perform a scaling analysis of the reconnection rate as a function of the opening angle made by the upstream magnetic fields, finding a maximum reconnection rate close to 0.2. The predictions compare favorably to particle-in-cell simulations of relativistic electron-positron and non-relativistic electron-proton reconnection. The fact that simulated reconnection rates are close to the predicted maximum suggests reconnection proceeds near the most efficient state allowed at the MHD-scale. The rate near the maximum is relatively insensitive to the opening angle, potentially explaining why reconnection has a similar fast rate in differing models. 

\end{abstract}


\maketitle

{\it Introduction--} Magnetic energy is abruptly released
in solar and stellar flares \cite{Masuda94,Gershberg05,klimchuk15a},
substorms in magnetotails of Earth and other planets
\cite{kepko15a,Vasyliunas83}, disruptions and the sawtooth crash in
magnetically confined fusion devices \cite{yamada94a}, laboratory
experiments \cite{Yamada06}, and numerous high energy astrophysical
systems \cite{sironi15a,zweibel09a}.
Magnetic reconnection, where a change in topology of the magnetic
field allows a rapid release of magnetic energy into thermal and
kinetic energy, is a likely cause.  The reconnection electric field
parallel to the X-line (where magnetic field lines break) not only
determines the rate that reconnection proceeds, but can also be
crucial for accelerating energetic super-thermal particles.  It was
estimated that a normalized reconnection rate of $\simeq$ 0.1 is
required to explain time scales of flares and substorms
\cite{Parker73}.

Reconnection rates have been studied observationally, experimentally,
theoretically, and numerically.  Measurements can be {\it in situ},
such as in the magnetosphere and lab, or remote, as in solar and
astrophysical contexts.  Reconnection rates from these different
vantage points can be the same but need not be; for example, flux
ropes in the corona have macroscopic forces that can influence the
evolution of current sheets where reconnection occurs.  Therefore, it
is important to distinguish between system scales.  We define {\it
  global-scale} as system-size scale of magnetic domains.  The {\it
  local-scale} is a smaller MHD-scale region where the magnetic field
and plasma parameters achieve relatively uniform conditions upstream
of the diffusion region.  The {\it micro-scale} is the scale of the
diffusion region.  Here we focus on reconnection rates at the local-
and micro-scales;
coupling to global scales is beyond the scope of this paper.

The original model for the {\it local} reconnection rate was the
Sweet-Parker model \cite{sweet58a,parker57a}, but it was too slow to
explain observed time scales of flares and substorms \cite{parker63a}.
The collisional diffusion region is long and thin ({\it i.e.,} the
upstream magnetic fields have a small opening angle), developing a
bottleneck that keeps the inflow speed small. The Petschek model
\cite{petschek64a} was much faster by producing an open outflow region
({\it i.e.,} a larger opening angle), but is not a self-consistent
model \cite{biskamp86a,sato79a}.

The collisionless limit is more appropriate for many systems of
interest.  Two-dimensional (2D) local simulations of isolated, thin,
Harris-type current sheets reveal that the steady-state reconnection
has a fast rate of 0.1 \cite{shay99a} when normalized by the magnetic
field and Alfv\'en speed at the local-scale.  This rate is independent
of simulated electron mass \cite{shay98a,hesse99a} and system size
\cite{shay99a,hesse99a}. In particular, the GEM challenge study
\cite{birn01a} showed that the rate is comparable in Hall-MHD, hybrid,
and particle-in-cell (PIC) simulations. Consequently, it was argued
that the Hall term, the minimal non-ideal-MHD term in all three
models, is the key physics for producing the fast rate
\cite{rogers01a,drake08a}. However, further studies have raised
important questions. One gets similar fast rates in electron-positron
plasmas, for which the Hall term vanishes
\cite{bessho05a,daughton07a,swisdak08a}, and in the strong
out-of-plane (guide) magnetic field regime
\cite{yhliu14a,tenbarge14a,stanier15a,cassak15a} for which the Hall
term is inactive.  Even within resistive-MHD, the same 0.1 rate arises
when a localized resistivity is employed \cite{sato79a}.  This
evidence calls into question whether the Hall term is the critical
effect.  It was suggested that the appearance of secondary islands
could provide a universal mechanism for limiting the length of the
diffusion region \cite{daughton07a,yhliu14a}, but this model is also
not satisfying since the same rate is obtained even when islands are
absent \cite{stanier15b,drake08a}.

{\it In situ} magnetospheric observations reveal (local) reconnection
rates near 0.1 \cite{Blanchard96,Wang15}.  Solar observations suggest
(global) reconnection rates can be this high as well
\cite{Qiu02,Lin05,Isobe05}, or somewhat lower
\cite{Ohyama98,Yokoyama01}.  Therefore, observations suggest the {\it
  local} rate is 0.1, and the {\it global} rate can be at or below
0.1.  This also has numerical support; in island coalescence, the {\it
  global} rate can be lower than 0.1
\cite{karimabadi11b,stanier15c,ng15a}, while the {\it local} rate
remains close to 0.1 \cite{karimabadi11b}.

What causes the {\it local} reconnection rate to be $\sim$ 0.1 across
different systems remains an open question [e.g., Ref.~\cite{comisso16a}].  
In this paper, we offer a new approach to this long-standing problem.  We propose that the {\it
  local} rate has a maximum as a result of constraints at MHD scales
(rather than physics at the diffusion-region-scale as is typically
discussed).  We perform a scaling analysis to derive the maximum {\it
  local} rate for low-$\beta$ plasmas, which we find is $O(0.1)$.  The
fact that local simulations produce rates close to this maximum value
suggests that steady reconnection proceeds at a rate nearly as fast as
possible.  We show the predictions are consistent with PIC simulations
of a relativistic electron-positron plasma and a non-relativistic
electron-proton plasma.

{ \it Simple model--} Let the thickness and length of the
(micro-scale) diffusion region be $\delta$ and $L$, respectively.  For
collisionless reconnection, $\delta$ is controlled by inertial or
gyro-radius scales \cite{shay04a}.  If the opening angle made by the
upstream magnetic field is small, the diffusion region is long and
thin. Reconnection in this case is very slow, as in Sweet-Parker
reconnection \cite{sweet58a,parker57a}.  As the opening angle
increases, reconnection becomes faster. This is true to a point, but
cannot continue for all angles for two reasons. First, in order to
satisfy force balance, the upstream region develops structures over a
larger scale, as in the classical Petschek-type analyses
\cite{petschek64a,Priest86}; this is what we define as the
local-scale.  Since the diffusion region thickness continues to be
controlled by micro-scales, the diffusion region becomes embedded in a
wider structure \cite{shay04a,cassak09a,yhliu15a} of local-scale
$\Delta z$, where the magnetic field and plasma parameters achieve
relatively uniform upstream conditions. The magnetic field $B_{xm}$
immediately upstream of the diffusion region becomes smaller than the
asymptotic magnetic field $B_{x0}$. (The subscript ``$0$'' indicates
asymptotic quantities at the {\it local-scale} and ``{\it m}''
indicates quantities at the {\it micro-scale}.)  This is crucial
because it is $B_{xm}$ that drives the outflow from the diffusion
region; as it becomes smaller, reconnection proceeds more slowly.

\begin{figure}
\includegraphics[width=8cm]{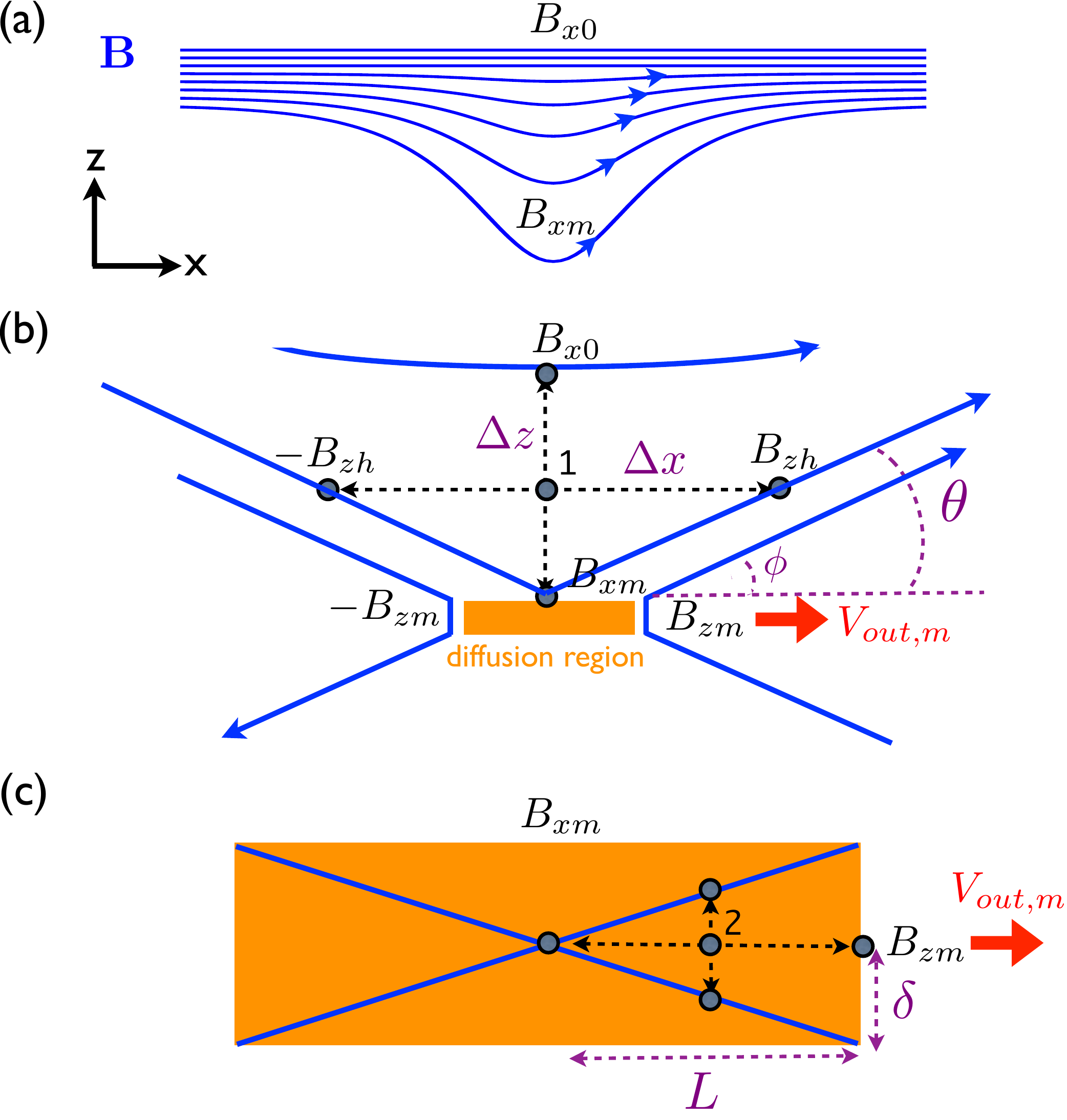}
\caption {(a) Sketch of magnetic field lines upstream of the diffusion
  region ($z >0$).  (b) Geometry of reconnection at the local-scale.  (c) Dimensions of
  the diffusion region at the micro-scale.}
\label{model}
\end{figure}

The second reason reconnection does not become faster without bound is
that the ${\bf J}\times {\bf B}$ force of the reconnected field
becomes smaller as the opening angle increases \cite{hesse09a}.
In the limit where the separatrices are
at a right angle, the tension force driving the outflow is canceled by the magnetic pressure force, so
reconnection does not spontaneously occur.

These observations suggest the following: the reconnection rate has a
maximal value for an intermediate opening angle which is large enough
to avoid the bottleneck for extremely thin current layers, but is not
too large to weaken the reconnection drive. 
We present a scaling analysis simply capturing these main aspects
using only the reconnection geometry and force balance.  We consider
low-$\beta$ systems in the relativistic limit; a more general
derivation should be future work.

The inflow region is illustrated in Fig.~\ref{model}(a).  With the
diffusion region at the micro-scale, the asymptotic (local) magnetic
field (at the top) must bend as it weakens toward the diffusion region
(at the bottom).  In the $\beta \ll 1$ limit, thermal pressure is
negligible, so to remain near equilibrium the inward-directed magnetic
pressure gradient force $-(\nabla B^2/8\pi)_z$ must be almost
perfectly balanced by outward-directed magnetic tension ${\bf
  B}\cdot\nabla B_z/4\pi$.  Evaluating these at point 1 marked in
Fig.~\ref{model}(b) gives
\begin{equation}
\frac{B_{x0}^2-B_{xm}^2}{8\pi\Delta z}\simeq
\left(\frac{B_{x0}+B_{xm}}{2}\right)\frac{2B_{zh}}{4\pi\Delta x},
\label{up_force}
\end{equation}
where $B_{zh}$ is evaluated at the upstream field line near the
separatrix.

We make the reasonable assumption the opening angle made by the
upstream field at the local-scale, $\theta\equiv\mbox{tan}^{-1}(\Delta z/\Delta
x)$, matches the opening angle of the micro-scale field at the corner
of the diffusion region, $\phi\equiv\tan^{-1}(B_{zm}/B_{xm})$.  Then, from
geometry, we get $B_{zh}/[(B_{x0}+B_{xm})/2] \simeq \Delta z/\Delta x
\simeq B_{zm}/B_{xm}$.  Eliminating $B_{zh}$ and solving for $B_{xm} /
B_{x0}$ gives
\begin{equation}
\frac{B_{xm}}{B_{x0}}\simeq \frac{1-(\Delta z/\Delta x)^2}{1+(\Delta
  z/\Delta x)^2}.
\label{BzL_BxL}
\end{equation} 
For small opening angles, $B_{xm} \simeq B_{x0}$; for large opening
angles approaching $45^\circ$, $B_{xm} \ll B_{x0}$, and embedding is
significant.

To estimate the outflow speed, we employ force balance in the
$x$-direction at point 2 in Fig.~\ref{model}(c).  In the relativistic
limit \cite{hesse07a}, $n'm_iU_{out}^2/2L+B_{zm}^2/8\pi L\simeq
(B_{zm}/2)2(B_{xm}/2)/4\pi\delta$, where $n^\prime$ is the density
measured in the fluid rest frame, $m_i$ is the ion mass, $U_{out}$ is
the $x$-component of the 4-velocity. Note that we have
  assumed that the profile of plasma pressure in the outflow direction
  is nearly uniform, as has been done in previous analyses
  \cite{birn10a}, so that the pressure gradient force is small compared
  to the magnetic tension force. The outflow speed $V_{out,m}$ from
the end of the diffusion region is related to $U_{out}$ through
$U_{out} = \gamma_{out}V_{out,m} =
V_{out,m}/(1-V_{out,m}^2/c^2)^{1/2}$, where $\gamma_{out}$ is the
relativistic factor.  Since the separatrix goes through the corner of
the diffusion region, $B_{zm}/B_{xm}\simeq \delta/L$. Solving for the
outflow speed as a function of $\delta/L$ gives
\begin{equation}
V_{out,m}\simeq
c\sqrt{\frac{(1-\delta^2/L^2)\sigma_{xm}}{1+(1-\delta^2/L^2)\sigma_{xm}}},
\label{vout}
\end{equation}
where the magnetization parameter evaluated near the diffusion region
is $\sigma_{xm}=B_{xm}^2/ 4\pi n' m_ic^2$. Consequently, if $\delta/L
\ll 1$, then $V_{out,m}\sim V_{Am}$ as expected since the Alfv\'en
speed in the relativistic limit \cite{sakai80a} is
$V_{Am}=c[\sigma_{xm}/(1+\sigma_{xm})]^{0.5}$.  However, as $\delta/L
\rightarrow 1$, the outflow speed $\rightarrow 0$ \cite{hesse09a}.

Putting the results together yields a prediction for the normalized
{\it local} rate.  The reconnection electric field $E_y$ is $B_{zm}
V_{out,m}/c$.  The reconnection rate $R_0 \equiv cE_y/B_{x0}V_{A0}$
normalized to local quantities is
\begin{equation}
  R_0\simeq
  \left(\frac{B_{zm}}{B_{xm}}\right)\left(\frac{B_{xm}}{B_{x0}}\right)
  \left(\frac{V_{out,m}}{V_{A0}}\right).
\label{RG}
\end{equation}
The rate normalized to the micro-scale magnetic field and Alfv\'en speed is $R_m\simeq
(B_{zm}/B_{xm})(V_{out,m}/V_{Am})$, and the micro-scale inflow speed is $V_{in,m}\simeq R_m V_{Am}$.

\begin{figure}
\includegraphics[width=7cm]{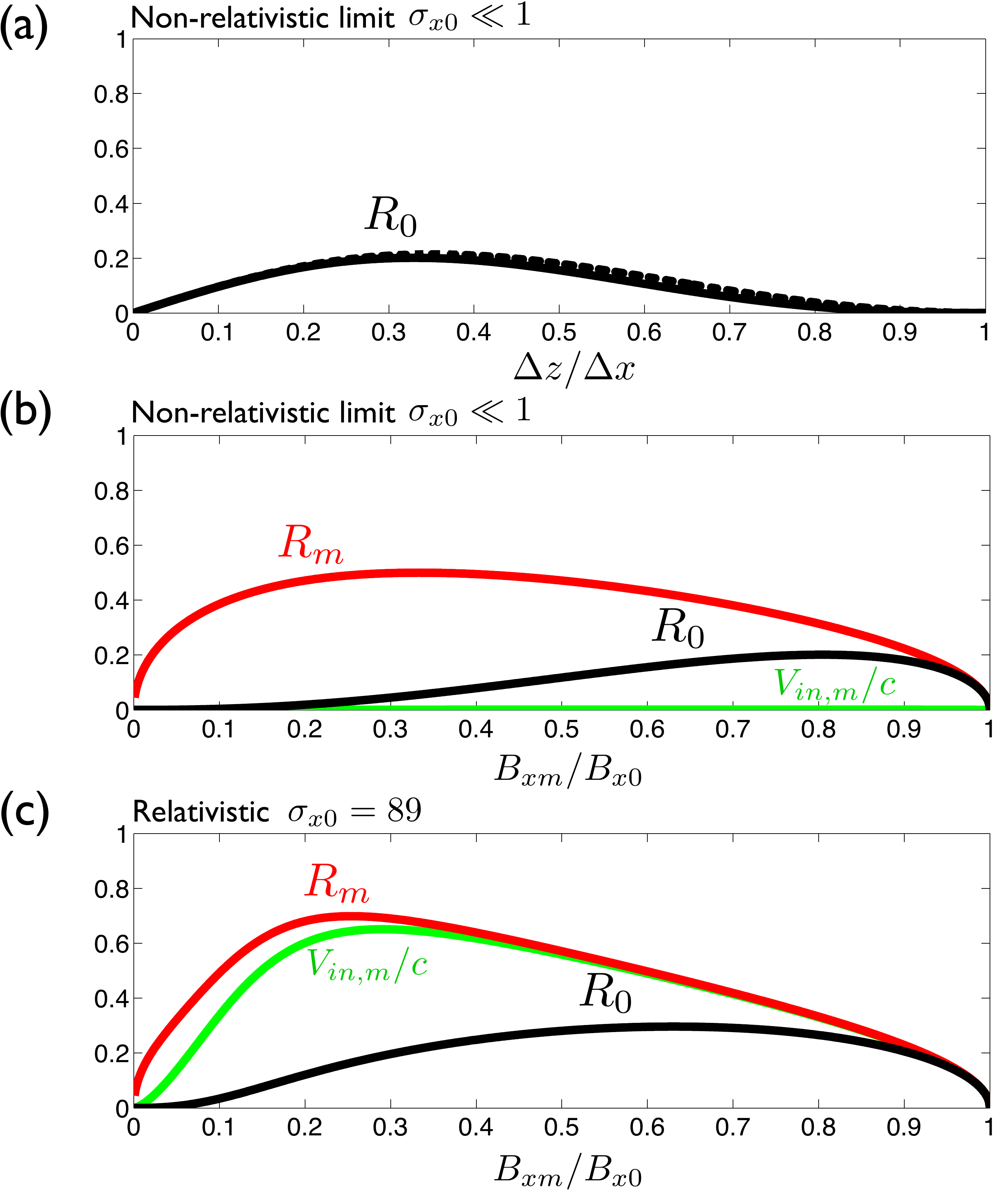}
\caption {Predictions for the non-relativistic limit as functions of
  (a) $\Delta z/\Delta x$ and (b) $B_{xm}/B_{x0}$.  (c) Predictions
  for relativistic limit ($\sigma_{x0}=89$).}
\label{prediction}
\end{figure}

Writing Eqs.~(\ref{BzL_BxL}) and (\ref{vout}) as functions of $\Delta
z/\Delta x$ and substituting into Eq.~(\ref{RG}) gives the predicted
{\it local} rate.  In the non-relativistic limit ($\sigma_{x0}\ll 1$),
the rate is
\begin{equation}
    R_{0,NR}\simeq \frac{\Delta z}{\Delta x}
  \left[\frac{1-(\Delta z/\Delta x)^2}{1+(\Delta z/\Delta x)^2} \right]^2
  \sqrt{1-\left(\frac{\Delta z}{\Delta x}\right)^2},
\end{equation}
which is plotted as the solid curve in Fig.~\ref{prediction}(a). This
expression generalizes the previously known result \cite{sweet58a,parker57a,hesse09a} of $R_{0,NR} \simeq
\Delta z /\Delta x \simeq \delta/L$ for small opening angles. In the $\Delta z/\Delta x \rightarrow 0$
and $\Delta z/\Delta x \rightarrow 1$ limits, $R_{0,NR}$
vanishes. Between the two extremes, $R_{0,NR}$ has a maximum,
conforming to the discussion earlier.  The maximum occurs at $\Delta
z/\Delta x\simeq0.31$ corresponding to a rate of 0.2, close to the
fast rate of order 0.1 widely observed. More importantly, the {\it
  local} rate is relatively flat for a broad range of $\Delta z/\Delta
x$ around the optimal value, suggesting that the rate is not strongly
sensitive to the opening angle for intermediate values.  This may
explain why reconnection rates in disparate physical systems are so
similar. 

\begin{figure}
\includegraphics[width=8cm]{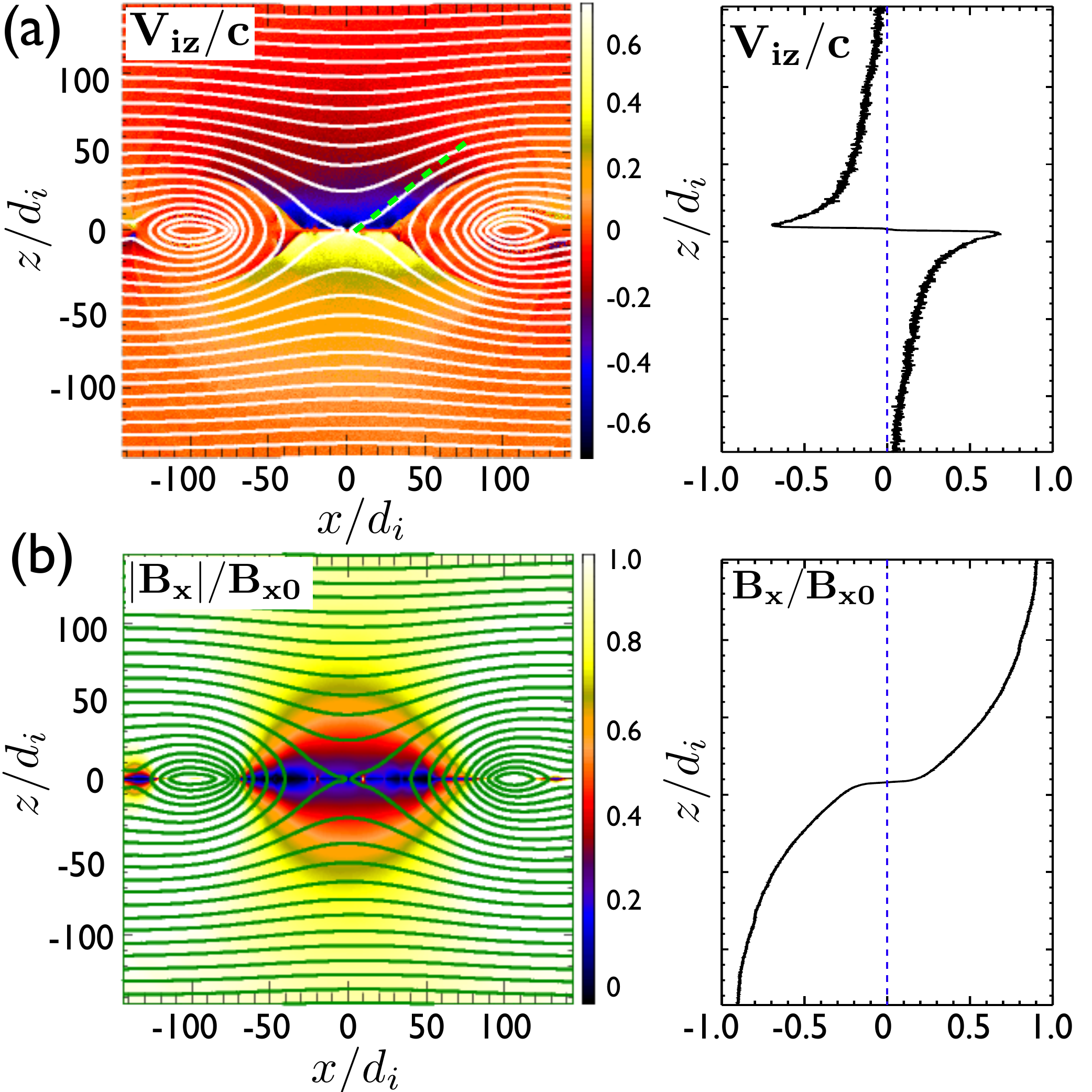} 
\caption {Local-scale structure around the X-line in electron-positron
  reconnection with $\sigma_{x0}=89$ at $t=600/\omega_{pi}$. (a)
  $V_{iz}$ and its cut at $x=0$.  (b) $|B_x|$ and a cut of $B_x$ at
  $x=0$.  Contours of in-plane magnetic flux are overlaid.  The color
  table in (b) has an upper limit $B_{x0}$.}
\label{feature}
\end{figure}

The dotted curve in
Fig.~\ref{prediction}(a) shows the non-relativistic prediction if
$V_{out,m}$ is taken to be identically $V_{Am}$ in Eq.~(\ref{RG}).
This comparison indicates that the correction to the outflow speed in
Eq.~(\ref{vout}) does not significantly alter $R_0$, although it does
impact $R_m$ as $\Delta z/\Delta x$ approaches 1.  Thus, the most
significant effect limiting the {\it local} rate with a increasing
opening angle is the embedding.
We plot $R_0$, $R_m$ and $V_{in,m}/c$ in Fig.~\ref{prediction}(b) as
functions of $B_{xm}/B_{x0}$ to facilitate a comparison with
simulations.  A similar plot is shown in Fig.~\ref{prediction}(c) for
the relativistic limit, specifically with $\sigma_{x0} = 89$.  The
peak $R_0$ is 0.3, and it does not change with increasing
$\sigma_{x0}$.  This bounds rates seen in relativistic simulations
\cite{sironi16a,yhliu15a,FGuo15a,bessho12a,melzani14a}.

  We point out that there are similarities between the
  present model and the classical Petschek model \cite{petschek64a}.
  However, there are a number of important differences.  For example,
  the Petschek model assumes a value of 0.5 for what we call
  $B_{xm}/B_{x0}$, whereas we estimate it self-consistently.
  Furthermore, the way Petschek obtained the upstream condition,
  strictly speaking, only works for the small opening angle limit,
  while our result is valid for any opening angle.  Finally, and most
  importantly, the weak dependence on reconnection rate reported by
  Petschek has a logarithmic dependence on Lundquist number, so the
  normalized reconnection rate is not bounded by 0 and 1 as it must be
  on physical grounds.  In the present work, the reconnection rate is
  manifestly bounded between 0 and 1.

{\it Comparison to particle-in-cell simulations--} We
compare the predictions against PIC simulations of a relativistic
electron-positron plasma ({\it i.e.,} $m_i=m_e$) in
Ref.~\cite{yhliu15a}.
The upstream magnetization parameter $\sigma_{x0}=89$ and
$\beta=0.005$.  The diffusion region is embedded, as is clearly seen
in Fig.~\ref{feature} which shows the inflow velocity $V_{iz}$ and
reconnecting magnetic field $B_x$ with in-plane magnetic flux overlaid
at time $t=600 / \omega_{pi}$. A vertical cut through the X-line of
these quantities is also shown. Immediately upstream of the diffusion
region of $d_i$-scale thickness, $|V_{iz}|$ peaks at $\simeq 0.65 c$
and $|B_x|/B_{x0}$ drops to $\simeq 0.2$ ($d_i = c / \omega_{pi}$ is
the ion inertial scale).  The variation of the magnetic structure
extends $\gtrsim 100 d_i$ upstream, and the separatrix has an opening
angle wider than typically seen in the non-relativistic regime with
$\beta\sim O(1)$ ({\it e.g.}, \cite{shay98a}).  It was shown in
Ref. \cite{yhliu15a} that the magnetic pressure gradient balances
magnetic tension in the upstream region, as expected for this
low-$\beta$ system.

\begin{figure}
\includegraphics[width=8.5cm]{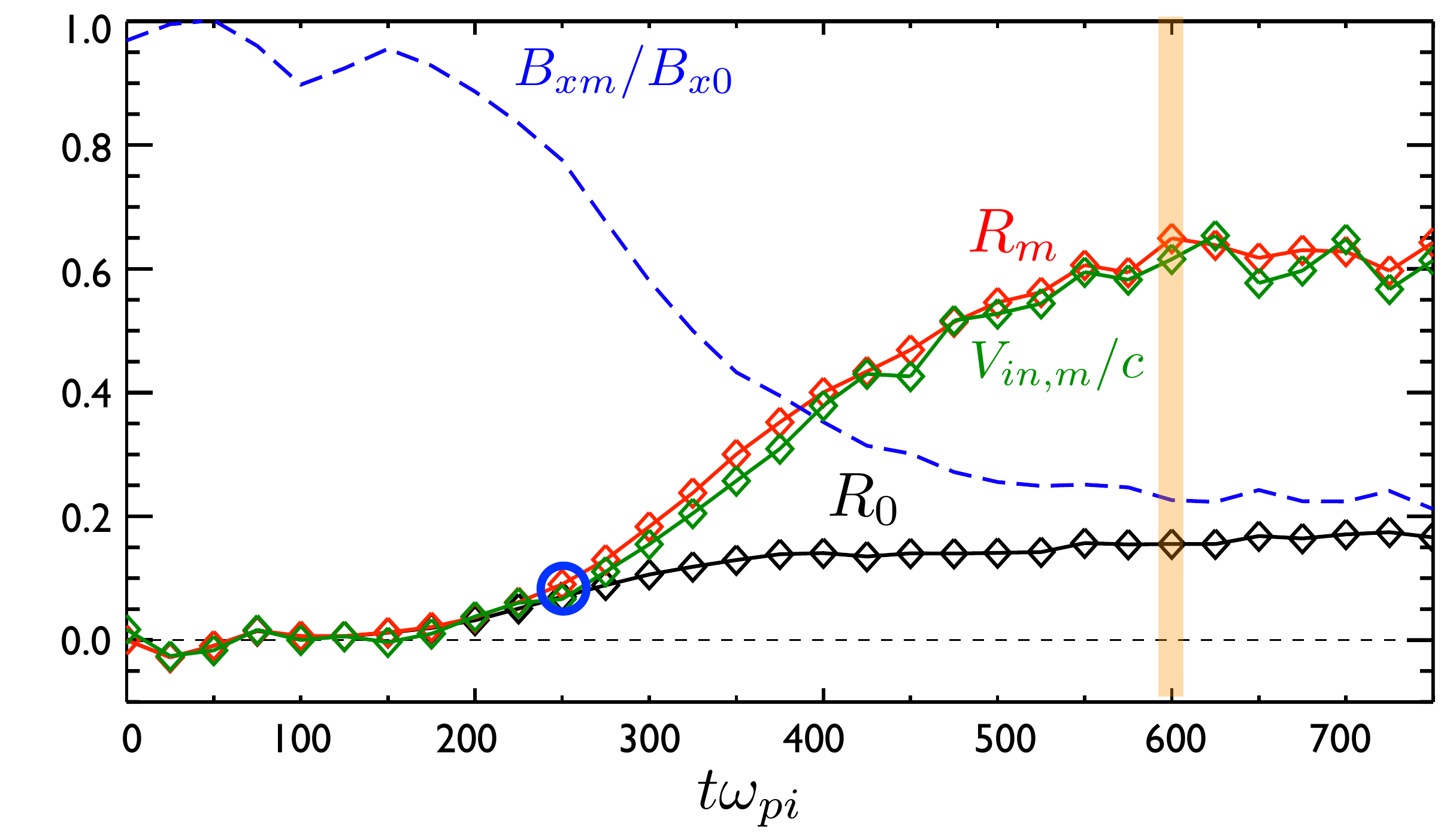} 
  \caption {Time evolution of the measured local reconnection rate
    $R_0$, micro-scale rate $R_m$, micro-scale inflow speed
    $V_{in,m}/c$ and $B_{xm}/B_{x0}$. The blue circle marks the
    deviation of $R_m$ from $R_0$. The orange vertical line marks the
    time plotted in Fig.~\ref{feature}.}
\label{rate}
\end{figure}

The time evolution of reconnection rates are
plotted in Fig.~\ref{rate}, along with the micro-scale inflow speed
$V_{in,m}$ and the ratio of magnetic fields $B_{xm}/B_{x0}$. Before a
quasi-steady state is reached, both $R_m$ and $R_0$ increase as the
simulation progresses. The deviation of $R_m$ from $R_0$ occurs at
time $t\simeq 250 / \omega_{pi}$ and $B_{xm}/B_{x0}\simeq 0.8$. $R_0$
reaches a plateau of $\simeq 0.15$ at $t \gtrsim 300/ \omega_{pi}$
while $R_m$ continues to grow and $B_{xm}/B_{x0}$ continues to drop. 
Note that, $R_0\simeq 0.15$ is reached before the generation of secondary tearing modes at $t\simeq 500 / \omega_{pi}$, indicating that $R_0$ is not determined by secondary islands.
$R_m$ eventually reaches a plateau of $\simeq 0.6$ and $B_{xm}/B_{x0}$
drops to $\simeq 0.22$ at $t\gtrsim 600 / \omega_{pi}$. The inflow
speed $V_{in,m}$ traces $R_m$ because $V_{Am}\simeq c$.  We compare
the steady-state values to the prediction shown in
Fig.~\ref{prediction}(c).  Substituting the measured
$B_{xm}/B_{x0}\simeq 0.22$ into the predictions gives $R_0 \simeq
0.14$, $R_m \simeq 0.69$, $V_{in,m} \simeq 0.62c$ and an opening angle
$\theta \simeq 38.7^\circ$ [illustrated by the dashed green line in
  Fig.~\ref{feature}(a)]. Given the simplicity of this model, this
agreement is quite remarkable.

To test the predictions against non-relativistic electron-proton
reconnection, we compare with a low-$\beta$ PIC simulation in
Ref.~\cite{PWu11a}, which used $m_i/m_e=25$ and a background density
$1\%$ of the sheet density. The upstream $\beta=0.01$.
From that reference, $R_0\simeq 0.085$ and $R_m \simeq 0.22$ with
$B_{xm}/B_{x0}\simeq 0.55$. The predictions in
Fig.~\ref{prediction}(b) based on $B_{xm}/B_{x0}\simeq 0.55$ are
roughly twice these values, mainly because of an overestimation of the
outflow speed. However, it is common for electron-proton plasmas to
have outflow speeds about half of the Alfv\'en speed \cite{PWu11a};
this is likely due to a self-generated firehose-sense temperature
anisotropy in reconnection exhausts, which reduces the outflow speed
({\it e.g.}, \cite{yhliu12a,yhliu11b}) but is not considered in the
current model.  Adjusting for this factor of two, the predictions
agree quite well with the simulations.  The predicted opening angle
$\sim 28.8^\circ$ purely based on the upstream constraint in
Eq.~(\ref{BzL_BxL}) agrees well.

{\it Discussion--} 
The model presented here is completely independent
of dissipation mechanism. The only ingredients are MHD-scale
considerations and that the diffusion region remains at micro-scales
when the opening angle increases. The fact that the simulated fast
rates in disparate physical models are all similar to the predicted
maximum rate of order 0.1 suggests that MHD-scale constraints on
magnetic energy release determine the fast rate.  The obvious
counterpoint to this is reconnection in MHD simulations with a uniform resistivity, which does not
proceed at this rate.  Even when the Lundquist number is large enough
to produce magnetic islands \cite{biskamp86a, loureiro07a}, the
reconnection rate is an order of magnitude smaller
\cite{daughton09a,YMHuang10a,shepherd10a}.  This indicates that
considerations at MHD scales are not sufficient to explain fast
reconnection; the micro-scale dissipation/localization mechanism must
be able to support the desired opening angle at the local-scale. 
However, if the diffusion region can support a larger
opening angle, the {\it local} rate of order $\sim O(0.1)$ is not strongly sensitive to the opening
angle over a wide range of values. The micro-scale rate $R_m$ is sensitive to the opening angle, resulting in the large difference between $R_m$ and $R_0$ observed in the relativistic limit \cite{yhliu15a}.

The present model is not complete in that it does not include some
physics that may affect the reconnection rate. As discussed earlier, 
the self-generated pressure anisotropy in the exhaust can
reduce the outflow speed \cite{yhliu12a,yhliu11b}.  
The plasma pressure gradient force in the outflow direction can also affect the outflow speed \cite{priest00a,yhliu14a,PWu11a}. 
Self-generated upstream temperature anisotropies \cite{egedal13a} may modify the
embedding.  Relaxing the low-$\beta$ assumption is important.
However, we note that the reduction of $B_{xm}$ also occurs in
simulations with $\beta\sim O(1)$ in Ref.~\cite{PWu11a}.  
Finally, this model does not take into account the conversion of upstream
energy into heat and accelerated particles, which undoubtedly impacts
the energy conversion process and is important in maintaining the
intense current sheet during reconnection \cite{hesse09a, hesse11a}. 
  Nevertheless, this simple model offers a new approach to
  the long-standing fast reconnection rate problem, which is broadly
  relevant in basic plasma physics, fusion science, solar and space
  physics, and astrophysics, and potentially provides an avenue for
  understanding the important link between the micro- and
  global-scales.  \\

\acknowledgments Y.-H. Liu thanks M. Swisdak and J. C. Dorelli for
helpful discussions, and P. Wu for sharing her simulation data. Y.-H. Liu is supported by NASA grant NNX16AG75G. M. Hesse acknowledges support by NASA's MMS mission. F. Guo is
supported by NASA grant NNH16AC601. H. Li is is supported by the DOE through the LDRD program at LANL and DOE/OFES support to LANL in collaboration with CMSO. P. A. Cassak acknowledges support
from NSF Grants AGS-0953463 and AGS-1460037 and NASA Grants NNX16AF75G
and NNX16AG76G. M. Shay is supported by NSF grant
AGS-1219382. Simulations were performed with LANL institutional
computing, NASA Advanced Supercomputing and NERSC Advanced
Supercomputing.\\

\end{document}